# Nontrivial doping evolution of electronic properties in Ising-superconducting alloys

*Wen Wan, Darshana Wickramaratne, Paul Dreher, Rishav Harsh, I. I. Mazin and Miguel M. Ugeda\**

W. Wan, P. Dreher, R. Harsh and M. M. Ugeda

Donostia International Physics Center (DIPC), Paseo Manuel de Lardizábal 4, 20018 San Sebastián, Spain
E-mail: mmugeda@dipc.org

D. Wickramaratne, Center for Computational Materials Science, U.S. Naval Research Laboratory, Washington, DC 20375, USA

I.I. Mazin, Department of Physics and Astronomy, George Mason University, Fairfax, VA 22030, USA

I.I. Mazin, Quantum Science and Engineering Center, George Mason University, Fairfax, VA 22030, USA

M. M. Ugeda, Centro de Física de Materiales (CSIC-UPV-EHU), Paseo Manuel de Lardizábal 5, 20018 San Sebastián, Spain.

M. M. Ugeda, Ikerbasque, Basque Foundation for Science, 48013 Bilbao, Spain.

*Transition metal dichalcogenides offer unprecedented versatility to engineer 2D materials with tailored properties to explore novel structural and electronic phase transitions. In this work, we present the atomic-scale evolution of the electronic ground state of a monolayer of $Nb_{1-\delta}Mo_{\delta}Se_2$ across the entire alloy composition range ($0 < \delta < 1$) using low-temperature (300 mK) scanning tunneling microscopy and spectroscopy (STM/STS). In particular, we investigate the atomic and electronic structure of this 2D alloy throughout the metal to semiconductor transition (monolayer $NbSe_2$ to $MoSe_2$). Our measurements let us extract the effective doping of Mo atoms, the bandgap evolution and the band shifts, which are monotonic with $\delta$. Furthermore, we demonstrate that collective electronic phases (charge density wave and superconductivity) are remarkably robust against disorder. We further show that the superconducting $T_C$ changes non-monotonically with doping. This contrasting behavior in the normal and superconducting state is explained using first-principles calculations. We show that Mo doping decreases the density of states at the Fermi level and the magnitude of pair-breaking spin fluctuations as a function of Mo content. Our results paint a detailed picture of the electronic structure evolution in 2D TMD alloys, which is of utmost relevance for future 2D materials design.*

# 1. Introduction

In transition metal dichalcogenides (TMD), the possibility of substituting one of the atomic elements by different species has greatly expanded the potential of this family of 2D materials. This route to obtain novel and stable TMD layers with tailored properties in the form of alloys ($M_{1-\delta}N_{\delta}X_2$ or $MX_{2(1-\delta)}Y_{2\delta}$ where M, N are TM and X, Y chalcogens, for binary alloys), has been recently demonstrated,[1,2] and shows significant promise in electronic,[3–7] optoelectronic[2,8–11] and catalytic[12–14] applications. So far, isovalent group VI TMD alloys such as $Mo_{\delta}W_{1-\delta}S_2$ and $WS_{2(1-\delta)}Se_{2\delta}$ have received most of the attention due to the possibility to tune the electronic and optical bandgaps.[2,8–11,15,16] Regarding aliovalent TMD alloys where TMs with different valences are involved, most efforts have focused on the dilute limit. Substitution of aliovalent species in TMD semiconductors at moderate concentrations enables stable modifications of the carrier type and density.[3,4,17–19] A paradigmatic example is the case of Nb:doped TMD semiconductors, where doping has been investigated in alloys up to 10% of Nb, and stable p-type conduction was demonstrated[3,4,6,7,17,20,21].

More fundamentally, TMD alloys are also an ideal playground for the study of 2D phase transitions and critical phenomena. Monolayer alloys that connect two $MX_2$ materials with different phases (structural, electronic, magnetic, etc.) should enable one to manipulate these phase transitions where the chemical composition of the alloy serves as the tuning parameter. However, the synthesis of a particular TMD alloy is subject to the miscibility of the components and its thermodynamic stability.[22] Most of the experimental progress towards this end was related to isovalent TMD alloys. For example, structural and electronic phase transitions have been studied in the isovalent $MSe_{2(1-\delta)}Te_{2\delta}$ with M = W, Mo.[23,24] Nonetheless, aliovalent alloys can bridge TMD materials with more disparate properties and, therefore, a richer variety of phase transitions can be accessed.[25,26] For instance, group-V TMD materials host collective phases such as charge density wave (CDW) order,[27–30] superconductivity (SC),[28,31–33] spin liquid behavior,[34] magnetism,[34–36] and topological phases.[37,38] These electronic phases are highly susceptible to external stimuli, and their study in TMD alloys enable access to, for example, relevant disorder and doping effects as well as their intrinsic robustness, which remain rarely explored,[24,39] in particular in the ultimate single-layer limit.

One of the latest and most intriguing additions to the collection of nontrivial collective states in the TMDs is a novel superconducting state dubbed "Ising superconductivity",[33] where spin-orbit coupling induced splitting of the singlet order parameter leads to a linear combination of momentum-separated singlet-triplet states, and an extraordinary protection against magnetic-field induced pair breaking. It has been proposed that impurity scattering plays a key role in Ising superconductivity, in particular, making the formally infinite in-plane critical field finite.[40]

In this work, we present an experimental and theoretical investigation of the aliovalent TMD alloy $Nb_{1-\delta}Mo_{\delta}Se_2$. We have successfully synthesized high-quality monolayers across the entire $0 < \delta < 1$ range, and examined the evolution of the atomic and electronic structure using low-temperature (0.3 – 4.2 K) STM/STS. First, our measurements enable us to explore the effect of electron doping on the monolayer $NbSe_2$ and track its impact on the electronic bands. The metal-semiconductor transition in this TMD alloy occurs for Nb concentrations of 25% ($\delta = 0.75$), which enables a wide-range across which the band gap can be tuned up to a maximum band gap of 2.2 eV of monolayer $MoSe_2$.[41] Second, we also observe a nearly simultaneous disappearance of the CDW and superconductivity, when the Mo concentration exceeds 15% ($\delta \gtrsim 0.15$), which highlights the robustness of these collective states against disorder.

A closer look reveals that the microscopic reasons for the destruction of CDW and SC are different. While the SC undergoes an initial strengthening at low Mo doping levels followed by a monotonic weakening, the CDW gradually loses the 3 x 3 long-range order. The effect of Mo on SC, including the initial increase, are due to doping and therefore a monotonic reduction of the density of states (DOS) at the Fermi level. The non-monotonic behavior appears due to the competition between two factors: the electron-phonon coupling strength is directly proportional to DOS, and the proximity to magnetism is resonantly related to the DOS. As a result, at low doping concentrations of Mo, the proximity to magnetism is reduced more rapidly than the coupling strength, ensuring the initial rise of $T_c$, but then becomes unimportant as the weakening of the electron-phonon interactions becomes the main factor. Instead, the weakening of the CDW order is weakly dependent on doping, and is mostly due to a systematic reduction of the CDW domain sizes due to Mo pinning the domain boundaries.

**2. Identification of substitutional transition metal dopants**

We have grown monolayers of the aliovalent alloy $Nb_{1-\delta}Mo_{\delta}Se_2$ with different compositions ($0 < \delta < 1$) on bilayer graphene (BLG)/6H-SiC(0001) by molecular beam epitaxy (MBE), as sketched in **Figure 1**a (see Methods for details). Graphene is both a suitable template for the synthesis of TMD materials and plays a negligible role in their electronic structure and collective electronic phases[42]. For dilute alloys ($\delta < 0.1$ and $\delta > 0.9$), STM imaging reveal that the element in minor concentration (Mo and Nb, respectively) is observed as individual point defects in the atomic lattice. This is illustrated in Figure 1b for $Nb_{0.93}Mo_{0.07}Se_2$ ($\delta = 0.07$). Figure 1b shows an atomically resolved STM image of the $NbSe_2$ lattice, where the 3x3 CDW ordering is also visible. In addition, several triangular features of equal orientation can also be observed. We attribute these features to Mo atoms located in substitutional positions in the Nb lattice. Since the STM images show the outermost chalcogen lattice, the triangular feature corresponds to the three Se atoms bonded to a single Mo substitutional atom underneath, whose local DOS is different from the Se atoms bonded to Nb atoms, thus link them visible. These triangular-like features are also predicted by our first-principles density functional theory (DFT) calculations of the electron density in the vacuum region above a Mo on the Nb site in NbSe$_2$, as illustrated in Figure 1d. The concentration of these triangular features is directly proportional to the Mo flux during the MBE growth (see SM), which further supports their identification as individual Mo atoms in the Nb lattice.

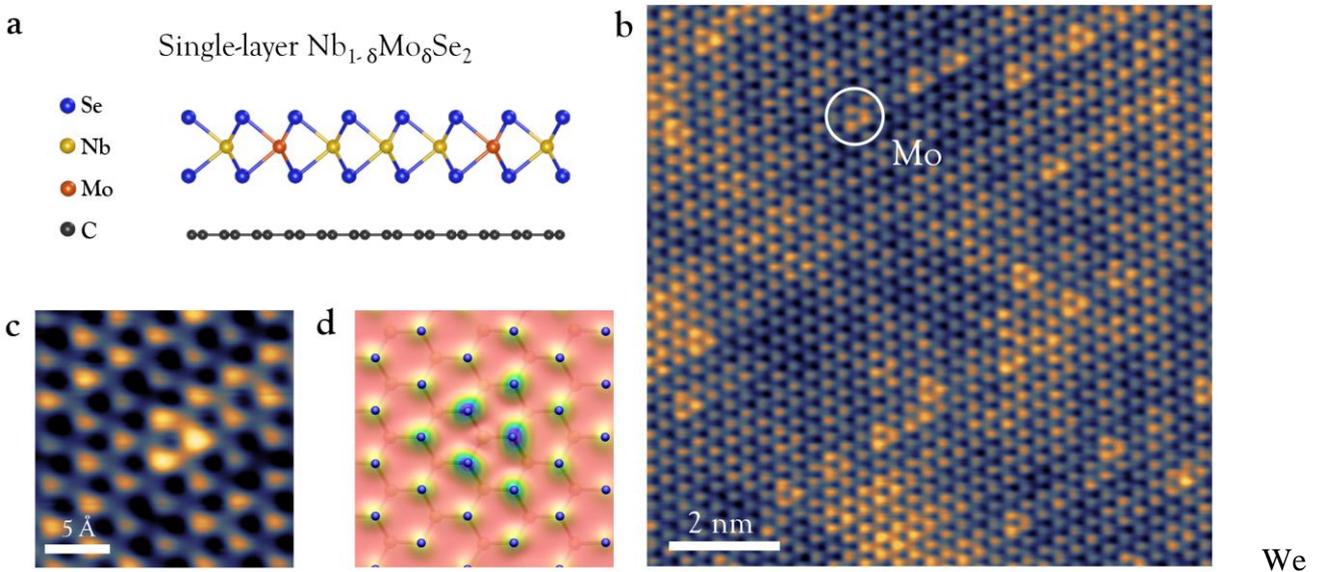

**Figure 1.** Atomic structure of the monolayer Nb$_{1-\delta}$Mo$_\delta$Se$_2$ alloy. a) Side view sketch of Nb$_{1-\delta}$Mo$_\delta$Se$_2$, including the graphene substrate. b) High-resolution STM image of Nb$_{0.02}$Mo$_{0.98}$Se$_2$ (V$_s$ = - 0.5 V, I$_t$ = 0.059 nA, T = 0.34 K). c) STM image of an individual substitutional Mo atom embedded in the NbSe$_2$ atomic lattice (V$_s$ = - 0.55 V, I$_t$ = 0.1 nA, T = 0.34 K). d) Charge density isosurface of the occupied Mo states for Mo substituting on the Nb site.

We have also analyzed the energetics of incorporating Mo in NbSe$_2$ and found it consistent with this interpretation. To this end, we have calculated formation energies for different possible configurations of Mo incorporated into NbSe$_2$: substituting on the Nb site (Mo$_{Nb}$), substituting on the Se site (Mo$_{Se}$), or being adsorbed on top of the NbSe$_2$ surface. For the latter, there are three possible configurations; Mo adsorbed right on top of a Nb, Mo$_{ads}^{Nb}$, or on top of a Se, Mo$_{ads}^{Se}$, or above the hollow site, Mo$_{ads}^{hollow}$. The formation energies for these configurations are summarized in **Table 1**. In the dilute limit Mo$_{Nb}$ has the lowest formation energy, which confirms our assertion that Mo indeed replaces Nb sites when incorporated into NbSe$_2$.

| Defect | Nb-rich (eV) | Nb-poor (eV) |
|---|---|---|
| Mo$_{Nb}$ | 1.01 | -0.14 |
| Mo$_{Se}$ | 3.41 | 6.12 |
| Mo$_{ads}^{hollow}$ | 2.60 | 4.03 |

**Table 1:** Formation energy of Mo under Nb-rich and Nb-poor conditions substituted on the niobium site, Mo$_{Nb}$, Mo substituted on the Se site, Mo$_{Se}$, and Mo adsorbed above the hollow site, Mo$_{ads}^{hollow}$, which is the most stable adsorption site for Mo on NbSe$_2$.

Our STM spectroscopy of substitutional Mo atoms (Mo$_{Nb}$) reveals two clear fingerprints in the low-lying electronic structure. **Figure 2**a shows a typical dI/dV curve (dI/dV ∝ LDOS) acquired on top of a triangular feature (yellow curve) along with a reference curve on pristine monolayer NbSe$_2$ (blue curve)[28]. The presence of this point defect induces a well-defined electronic resonance at 0.45 V below E$_F$ (black arrow). Differential conductance maps taken at this energy with a dilute concentration of Mo (Figure 2b) reveal that this resonance has a spatial extent of ~ 5Å around the defect.

The crystal field splitting of Mo$_{Nb}$ levels provides a key to understand the origin of this resonance. The Mo atom is in a trigonal prismatic coordination with Se when incorporated as Mo$_{Nb}$, which splits the five Mo *4d* states into three different groups as illustrated in Figure 2c. The density of states of Mo$_{Nb}$ from our DFT calculations highlights the Mo *d*-states where we find Mo $d_{z^2}$ states ($a_{1g}$ symmetry) at a lower energy compared to the Mo $d_{xy}, d_{x^2-y^2}$ states (*e'* symmetry), which is consistent with a trigonal crystal field acting on Mo$^{4+}$. The wavefunction of the Mo $d_{z^2}$ states, which extends out of the NbSe$_2$ plane, and are even within the *xy* plane, will penetrate into the vacuum region more compared to the planar and odd Mo $d_{xy}, d_{x^2-y^2}$ states and are therefore the most likely to contribute to tunneling.[43] In agreement with the energy of the resonance in the experiment, we find the Mo $d_{z^2}$ states to be located 0.47 eV below the NbSe$_2$ Fermi level in our calculations.

In addition to the resonance, Mo$_{Nb}$ also induces a rigid shift (grey arrows) of the peak labelled as E$_2$, which corresponds to the flat region at the top of the Nb-derived band at $\bar{\Gamma}$ of monolayer NbSe$_2$[28,42]. We will discuss this shift in more detail in **Section 4** and show that it is due to Mo$_{Nb}$ acting as a source of electron doping.

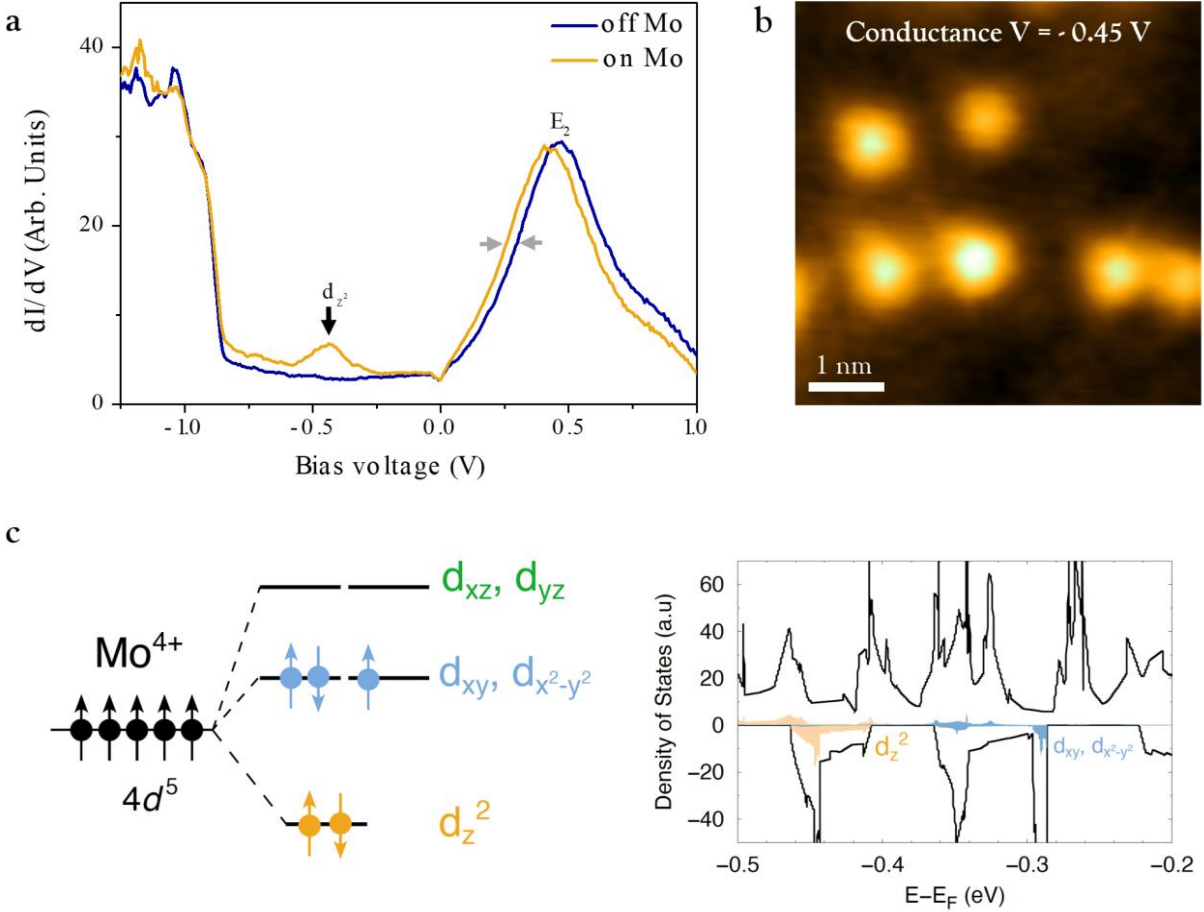

**Figure 2.** Electronic fingerprint of Mo substitutional atoms in monolayer NbSe$_2$. a) Wide-bias STM dI/dV spectra consecutively acquired on single-layer Nb$_{0.007}$Mo$_{0.993}$Se$_2$ on top of a substitutional Mo atom (yellow curve) and on a bare NbSe$_2$ region (blue curve). The grey arrows denote the shift in the conductance that is measured on top of the Mo atom versus off of the Mo atom. b) Experimental conductance maps taken in monolayer Nb$_{0.02}$Mo$_{0.98}$Se$_2$ taken at V$_s$ = - 0.45 V, the maximum of the d$_{z^2}$ states (I$_t$ = 2 nA, T = 4.2 K). c) Trigonal crystal field acting on the *4d*-states of Mo$^{4+}$ (left) and total spin-polarized density of states of Mo in a supercell of NbSe$_2$ (black) illustrating the Mo $d_{z^2}$ states (orange) and Mo $d_{xy}$, $d_{x^2-y^2}$ states (blue).

## 3. Atomic structure of the monolayer TMD alloys

Now that we have established the trigonal prismatic metal coordination of both TM elements by Se atoms in the TMD alloy, we characterize the morphology of the MBE-grown monolayer alloys as a function of δ. In total, we grew up to 24 TMD alloys to explore the whole range from NbSe$_2$ (δ = 0) to MoSe$_2$ (δ = 1). The composition of the grown TMD alloys relies on a calibration carried out in the dilute alloys (δ < 0.12 and δ > 0.9) by counting the number of individual dopants per area from high-resolution STM images (see SM for details). **Figure 3** shows the evolution of the morphology of the TMD alloys with δ through six STM images. Figure 3a and 3f show the topography of the pristine TMD monolayers, *i.e.,* NbSe$_2$ and MoSe$_2$, respectively. While both layers exhibit a high crystalline quality with a small density of point defects (< 10$^{12}$ cm$^{-2}$), monolayer MoSe$_2$ exhibits its characteristic mirror twin boundaries (MTB) forming triangular nanostructures.[44] For dilute concentrations of Mo (Figure 3b) and Nb (Figure 3e), individual dopants can be observed embedded within the host lattice, which still preserve their most characteristic features such as the 3x3 CDW order and the presence of MTBs, respectively. For higher concentrations (Figure 3c and 3d), dopants' electronic states overlap and the STM images are entirely dominated by the electronic structure.

STM imaging also allowed us to gain key information regarding the spatial distribution of the dopants across the monolayer. Our large-scale topographic images for δ < 0.5 reveal a homogeneous distribution of Mo atoms except very close to the edges (~10 nm), where a sudden depletion of Mo is observed (see **Figure S3**). This behavior at the edges was not noticeable at δ > 0.5. Although segregation of the atomic species seems minor at large length scales, the atomic distribution of Mo dopants in the monolayer shows evidence for atomic ordering. STM images of NbSe$_2$ with relatively high Mo concentration (δ > 0.15) reveal nm-sized stripes along the three crystallographic axes that likely represent Mo atomic chains (see Figure 3c and **Figure S4**). Although non-random dopant distributions are common in TMD alloys,[11,45,46] striped configurations usually exhibit strong anisotropy (parallel stripes only along one crystallographic direction), which affects their most fundamental properties.[11,46]

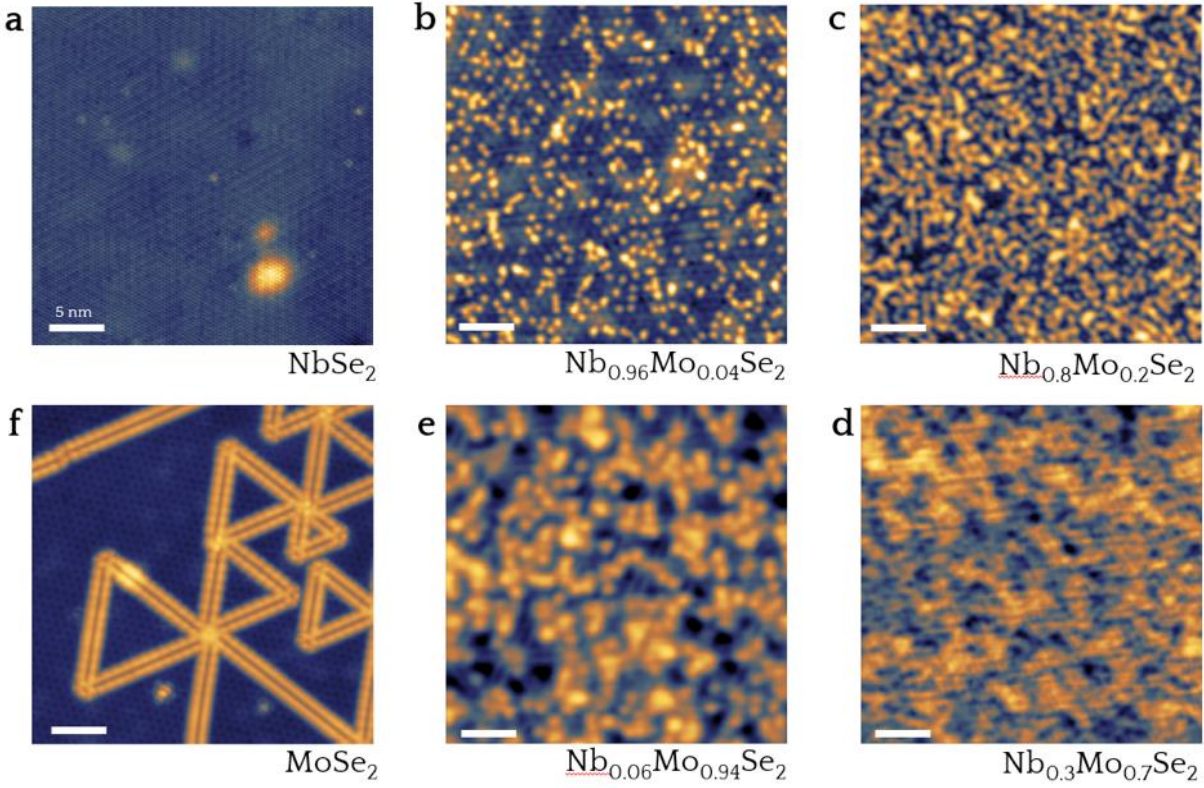

**Figure 3**. Atomic-scale morphology evolution of single-layer $Nb_{1-\delta}Mo_\delta Se_2$. a-f) High resolution STM images of several $Nb_{1-\delta}Mo_\delta Se_2$ monolayers for different $\delta$. All the images have the same size for an easier comparison (30 x 30 nm$^2$). a) $V_s$ = - 0.06 V, $I_t$ = 0.58 nA, T = 0.34 K. b) $V_s$ = - 0.55 V, $I_t$ = 0.2 nA, T = 0.34 K. c) $V_s$ = - 0.5 V, $I_t$ = 0.15 nA, T = 1.2 K. d) $V_s$ = - 0.8 V, $I_t$ = 0.01 nA, T = 4.2 K. e) $V_s$ = - 1.5 V, $I_t$ = 0.02 nA, T = 2.2 K. f) $V_s$ = - 1.5 V, $I_t$ = 0.05 nA, T = 4.2 K.

## 4. Evolution of the electronic structure: metal to semiconductor transition

We experimentally determined the evolution of the electronic structure of monolayers of $Nb_{1-\delta}Mo_\delta Se_2$ from the metallic $NbSe_2$ ($\delta$ = 0) to the semiconducting $MoSe_2$ ($\delta$ = 1). **Figure 4**a shows this evolution through a representative series of dI/dV spectra taken over a large bias range in steps of $\Delta\delta$ = 0.1. Each spectrum represents the averaged electronic structure of alloy areas in the range of 300–900 nm$^2$ sampled measuring grids of point dI/dV curves with ~2 curve/nm$^2$. Starting from $NbSe_2$ (bottom black curve), we find three characteristic STS features labelled as $E_i$, which are known to represent the onset of the three sets of electronic bands near E$_F$,[28] and which we can identify within our first-principles calculations. The $E_1$ feature marks the onset of the Se-$p$ band located at 0.81 V below E$_F$. The $E_2$ feature is a pronounced peak that corresponds to the onset of the Nb-d valence band at $\Gamma$ ($E_2$ = 0.46 V) with Nb $d_{z^2}$ character. This band is responsible for the metallic character of SL-$NbSe_2$ as well as the existence of both CDW and superconductivity. Lastly, the $E_3$ feature is believed to be the onset of the Nb conduction band, with a minimum at K. It is located approximately 2.1 eV above $E_F$, and has nearly pure $d_{z^2}$ character (in order to confirm these conjectures we have calculated DFT charge density profiles in vacuum above the NbSe$_2$ monolayer (**Figure S6**) and found that, indeed, the tunneling from the top of the valence band, which gives rise to $E_2$, is exponentially dominated by the $d_{z^2}$ states near Γ, while that from the conduction band, which gives rise to $E_3$, is dominated by the $d_{z^2}$ states near K, apparently due to this orbital extending farther in the z direction). The energy position of these key STS features can be tracked with the Mo concentration, as shown in Figure 4b. Interestingly, each of the features exhibits a unique shift (or lack thereof) as a function of δ. $E_1$ remains constant up to δ = 0.2. Unlike $E_1$, $E_2$ and $E_3$ exhibit a monotonic shift towards $E_F$ with δ.

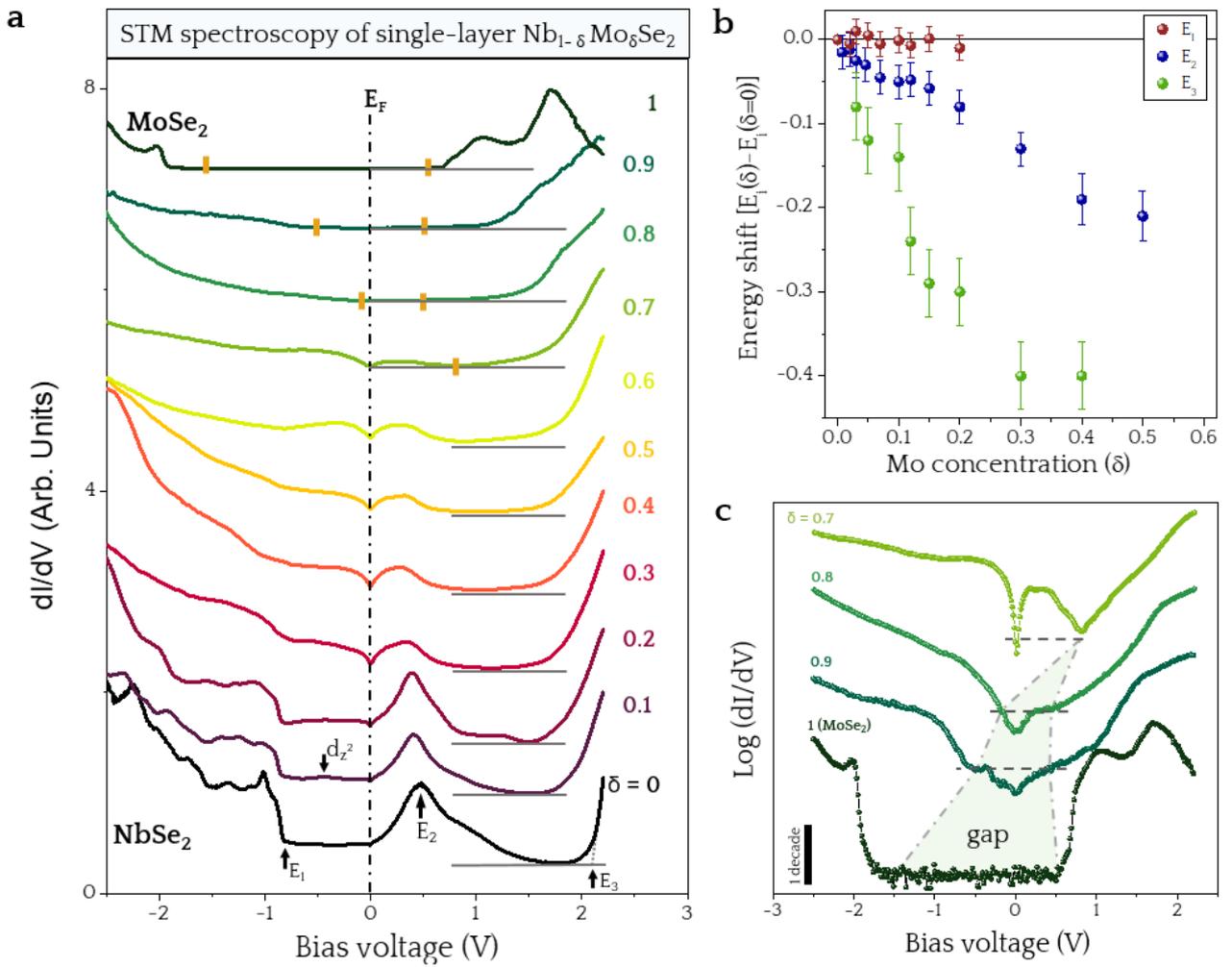

**Figure 4.** Electronic structure evolution of the monolayer Nb$_{1-\delta}$Mo$_\delta$Se$_2$ alloy. a) Representative wide-bias dI/dV spectra taken on different Nb$_{1-\delta}$Mo$_\delta$Se$_2$ monolayers across the full range 0 < δ < 1. Each curve represents the averaged electronic structure of alloy areas of 300 – 900 nm$^2$ sampled measuring grids of point dI/dV curves with ~ 2 curve/nm$^2$. b) Energy shift of the $E_{i=1-3}$ features with respect to their initial energy position in NbSe$_2$ (δ = 0) as a function of the Mo concentration (δ). c) Logarithmic plot of the dI/dV spectra relevant for the metal-semiconductor transition (0.7 < δ < 1).

To provide insight into the changes in $E_1$, $E_2$ and $E_3$ as a function of δ we consider different factors that affect the evolution of the electronic structure with Mo doping. This includes (i) electron doping since $Mo^{4+}$ has one more valence electron than $Nb^{4+}$, (ii) changes in *p-d* hybridization due to differences in the Mo and Nb ionic radii and (iii) changes in the dielectric properties as the electronic structure evolves from a metal to a semiconductor. The first effect shifts $E_F$ upwards. The second reduces the *p-d* separation, and therefore the separation between the corresponding bands. Indeed, the Nb bands move up by roughly $t_{pd}^2/(E_d - E_p)$, where $t_{pd}$ is some effective hybridization strength, and the Se bands move down by the same amount. Finally, in the semiconducting regime (not relevant for Figure 4b) the Coulomb interactions become long-range and density functional calculations need to be corrected (*e.g.*, with the use of hybrid functionals).

We first address the shift in $E_2$. This can be traced to the upward shift of the Fermi level mentioned above. This is indeed what we find with our first-principles calculations illustrated in **Figure 5**a. Doping $NbSe_2$ with Mo up to δ = 0.2, we find that $E_F$ shifts up by 0.071 eV with respect to pristine $NbSe_2$, in agreement with our experiment. For large Mo concentrations (δ > 0.5), the $E_2$ peak amplitude is largely damped, likely due to the significant broadening of the *d*-band. This coincides with the emergence of a significant DOS for energies below $E_F$. This peak in the DOS is the origin of the valence band of Nb-doped $MoSe_2$ monolayer, which develops from the localized $d_{z^2}$ state shown in Figure 2.

Next, we address why $E_1$ remains constant as a function of δ. As mentioned, the energy separation (which we define as $\Delta_\Gamma$) between the highest occupied transition metal *d*-state at Γ and the next lower lying state (which gives rise to the feature $E_1$) that is comprised of Se *p*-orbitals with admixture of transition metal $d_{xz}$, $d_{yz}$ orbitals is larger for $NbSe_2$ compared to $MoSe_2$. This is accompanied by a reduction in bandwidth of the valence band around Γ when going from $NbSe_2$ to $MoSe_2$ (see **Figure S5**a). Since the *d-d* direct overlap is small, the effective *d-d* hopping is controlled by the same parameter $t_{pd}^2/(E_d - E_p)$, albeit maybe with a different numerical coefficient. Hence, the reduction in $\Delta_\Gamma$ and in turn the bandwidth can be understood by a general reduction of $t_{pd}$, as a function of δ. Within this line of reasoning, we expect $\Delta_\Gamma$ to decrease as a function of δ, which is indeed what we find with our first-principles calculations (see Figure S5b). This upward shift of the Se derived band with respect to the highest occupied state at K as a function of δ coincides with the upward shift of the Fermi level due to doping (feature $E_2$). The cancellation of these two upward shifts results in an approximately constant energy difference between the Se *p*-states and the Fermi level, which is indeed what we find in our measurements of $E_1$ as a function of Mo concentration.

Feature $E_3$ is probably the most complex of all. At first glance, it reflects the crystal field gap, as it represents transitions between the Fermi level and the conduction $d$-band with the minimum at K. However, this state at K has pure $d_{z^2}$ character, similar to the band below at Γ. Therefore, the magnitude of the crystal field splitting is affected more by the dispersion of the $d_{z^2}$ band and $d_{z^2}/(d_{xy}, d_{x^2-y^2})$ hybridization. Since the $d$-$d$ hopping largely proceeds via the Se orbitals, it is again controlled by $t_{pd}^2/(E_d - E_p)$. Hence, within the simplest approximation the crystal field splitting should also decrease linearly as a function of δ, albeit with a different prefactor (note that in all three cases somewhat different combinations of $t_{pd\sigma}$ and $t_{pd\pi}$ form the effective $t_{pd}$). Indeed, in our experiments we find $E_3$ moves down roughly proportionally to, but faster than $E_2$. This is corroborated by our calculations (see Figure S5b), where we find the minimum crystal field gap in NbSe$_2$ is 1.33 eV, and in MoSe$_2$ it is 1.08 eV (both values are obtained with DFT, so for MoSe$_2$ it represents the gap in MoSe$_2$ with artificial metallic screening - which is what we need for the comparison with the experiment at δ < 0.6, where the material is still metallic). The difference is 0.25 eV. Experimentally, the $E_3$ feature flattens out with Mo doping at 0.4 eV, larger than, but qualitatively consistent with the DFT estimate.

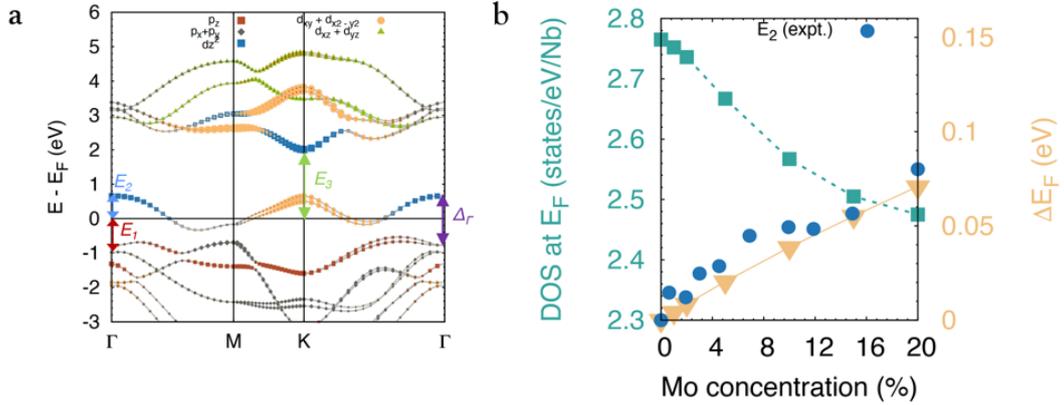

**Figure 5.** First-principles calculations of Mo doping in NbSe$_2$. a) Band structure of monolayer NbSe$_2$ plotted with respect to the Fermi level. The orbitals contributing to each state (listed in the legend in the top) are projected on to each band. The origin of $E_1$, $E_2$ and $E_3$ in our STM measurements (Figure 4) are marked with vertical arrows. b) Shift in the Fermi level (right vertical axis, orange triangle) and the magnitude of the density of states at the Fermi level (left vertical axis, teal squares) as a function of Mo concentration. Shift in the position of the peak labeled $E_2$ from the STM measurements (right vertical axis, blue circles) as a function of Mo content.

The electronic structure of these alloys as they evolve from NbSe$_2$ to MoSe$_2$ unavoidably undergoes a metal-semiconductor phase transition. This transition becomes visible at δ > 0.7, as shown in Figure 4a and Figure 4c. For δ = 0.7 (Nb$_{0.3}$Mo$_{0.7}$Se$_2$), while the alloy reaches zero DOS at E$_F$ and ~ 0.8 V, the gap is still negligible due to states originally from the *d*-band (E$_2$ feature). For δ = 0.8, these states disappear and a phase transition takes place, which leads to semiconducting behavior with an incipient p-type bandgap of 0.6 ± 0.2 eV. As δ increases up to MoSe$_2$, the bandgap grows due to the gradual depletion of the occupied states until it reaches the MoSe$_2$ (δ = 1) bandgap of 2.2 eV which is n-type.[41] Interestingly, the bandgap evolves across a broad range of Nb-doping of nearly 30%, which enables a large tunability.

## 5. Collective electronic states and disorder

Lastly, we experimentally investigate the properties of the superconducting and CDW orders subject to disorder in the TM plane, which is key to understand their development in NbSe$_2$. Similar to the metal-semiconductor transition, the evolution of the electronic structure of SL-Nb$_{1-δ}$Mo$_δ$Se$_2$ from SL-NbSe$_2$ to SL-MoSe$_2$ implies the occurrence of two electronic phase transitions, *i.e.*, the SC and CDW transitions.

First, we focus on the superconducting state. **Figure 6**a shows the evolution of the averaged SC gap with Mo concentration, which comprises two regimes (see SI for details in the SC fits). First, the SC gap undergoes a moderate increase of ~20% from that of SL-NbSe$_2$ (Δ = 0.40 meV) up to Δ = 0.48 meV for a Mo concentration of 3% (δ = 0.03). After that, the SC gap decreases monotonically and ultimately vanishes at Mo concentrations larger than 15% (δ = 0.15). Such a large value is remarkable and highlights the robustness of the SC state against non-magnetic structural disorder.

Based on the DOS of monolayer NbSe$_2$,[47] we expect doping NbSe$_2$ with electrons to decrease the DOS at the Fermi level, *N(E$_F$)*, compared to pristine NbSe$_2$. Our DFT calculations in Figure 6b confirm this, showing that at about δ = 0.2 doping, N(E$_F$) decreases by 15% compared to δ = 0. We expect this reduction in *N(E$_F$)* with doping to also impact spin fluctuations, which have been shown to be prominent in NbSe$_2$[47–49] and act as a source of pair-breaking.[50] We gauge the tendency to magnetism as a function of doping by performing fixed-spin moment (FSM) calculations of monolayer NbSe$_2$ doped with Mo concentrations of δ = 0-0.2 (Figure 6b). To this end, we used the VCA (see *Methods*) to determine the ferromagnetic spin susceptibility, χ, defined as $\chi = \left(\frac{\partial^2 E}{\partial m^2}\right)^{-1}$. We find that χ decreases monotonically as a function of increasing Mo content compared to pristine NbSe$_2$ (inset of Figure 6b), which implies the tendency to
magnetism decreases with increasing doping. Taken together we find that doping NbSe$_2$ with Mo leads to a reduction in *N(E$_F$)* and a reduction in the proximity to magnetism.

This behavior of $N(E_F)$ and χ as a function of δ yields important insight into the non-monotonic change in $T_c$. In the first approximation, the electron-phonon coupling constant, $λ_p$, is proportional to $N(E_F)$, while the electron-paramagnon coupling, $λ_s$, is proportional to the spin susceptibility $χ(q)$, averaged over the Brillouin zone. Using $χ(q = 0)$ as a proxy, it is proportional, in the standard random phase approximation, to $N(E_F)/[1-IN(E_F)]$, where the Stoner parameter $I$ is roughly independent of δ. In monolayer NbSe$_2$, $[IN(E_F)]$, was estimated to be rather large, ~0.7,[47] so that at small doping levels, $λ_s$ decreases due to the reduction in $N(E_F)$ three times faster than $λ_p$. At the BCS level, their difference $λ_p - λ_s$ enters in the equations for $T_C$.[51] Initially, the reduction of $λ_s$ is more important and drives $T_c$ up. Then both constants, $λ_p$ and $λ_s$, become too small and superconductivity wanes and eventually disappears. As a result of this competition, the actual dependence of $T_c$ on δ is non-monotonic (Figure 6a).

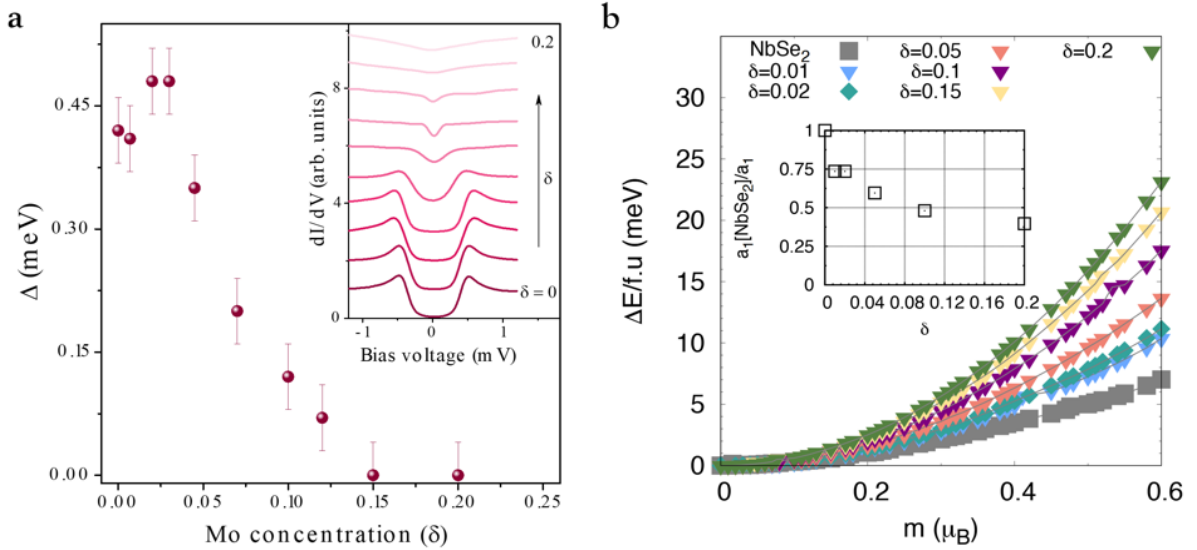

**Figure 6.** Superconductivity in the monolayer Nb$_{1-δ}$Mo$_δ$Se$_2$ alloy. a) Evolution of the experimental SC gap Δ with Mo concentration (δ). The inset shows the experimental dI/dV spectra from which the Δ values are extracted. Each spectrum represents the average over regions of typically ~ 100 nm$^2$. b) Collinear fixed-spin moment calculations of NbSe$_2$ as a function of doping with Mo illustrates the change in energy per formula unit with respect to the non-magnetic state as a function of magnetic moment per Nb atom. The inset illustrates the coefficient a$_1$ of NbSe$_2$ (see text) normalized by the value of a$_1$ obtained as a function of Mo content.

We now analyze the evolution of the CDW order with Mo content. **Figure 7**a shows a representative series of STM images for different δ values along with their corresponding FFT images in Figure 7b. As seen for the lowest Mo concentration (δ = 0.007), the 3x3 spots in the FFT are sharp with an intensity comparable to that of the (1x1) Bragg peaks, which indicates a fully developed CDW.

However, as the Mo concentration increases, the 3x3 peaks gradually broaden, become fainter and, ultimately, turn indistinguishable for δ ~ 0.2, a critical value for the CDW melting. This scenario is quantitatively reproduced in the plots in Figure 7c and 7d that show the averaged normalized peak intensity in the FFT ($I_{3x3}/I_{1x1}$) and the full width at half maximum ($w_{FWHM}$) of the 3x3 peaks.

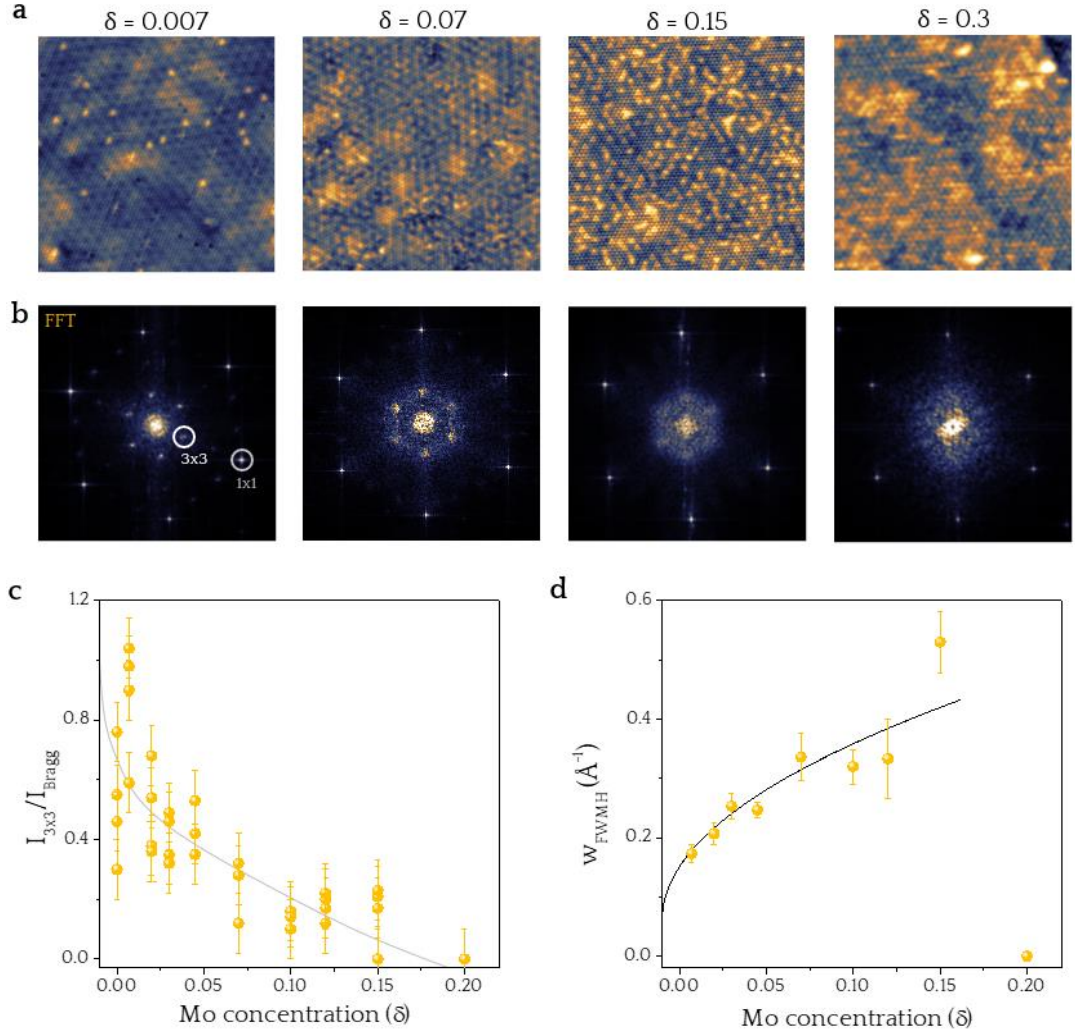

**Figure 7.** CDW in monolayer $Nb_{1-\delta}Mo_\delta Se_2$. a) High-resolution STM images for different Mo concentrations. Parameters: δ = 0.007 ($V_s$ = - 0.7 V, $I_t$ = 1.5 nA, T = 1.2 K), δ = 0.07 ($V_s$ = - 0.09 V, $I_t$ = 2 nA, T = 0.34 K), δ = 0.15 ($V_s$ = - 0.5 V, $I_t$ = 0.67 nA, T = 1.2 K) and δ = 0.3 ($V_s$ = + 0.02 V, $I_t$ = 1 nA, T = 2 K). b) Fast Fourier transform (FFT) of the STM images in a). c) Normalized intensity of the 3x3 peaks in the FFT of various experimental (yellow) atomically resolved STM images. The grey curve is a guide to the eye. d) Evolution of the width of the 3x3 peaks in the FFT as a function of Mo concentration. In black, the theoretical fit (see text).

One can anticipate two processes that contribute to the destruction of the CDW. First, as discussed, Mo impurities act as electron donors, and eventually turn the material semiconducting (at which point we do not anticipate a CDW to remain stable in pristine MoSe$_2$). Second, Mo impurities, as most defects, act as pinning centers for the phase domain walls.[52,53] Because of the latter effect, the CDW phase is disrupted at a length scale $l \sim 1/\sqrt{\delta}$, where $l$ is the average distance between two Mo impurities. At $\delta \approx 0.2$, $l \approx 2.2a$, and therefore a 3x3 CDW simply does not have room to develop. Nevertheless, as Figure 7b illustrates, even at $\delta \approx 0.3$ there still remains an incoherent cloud at about 1/3 r.l.u. from Γ, indicating that, while an ordered CDW is not possible any more, correlations of ionic displacement still "remember" the tendency to form the CDW with a **q** such that $|q| \approx 1/3$.

This latter observation attests to the fact that the melting of the CDW at $\delta \sim 0.2$ is due to the disorder introduced by Mo, and not due to electron doping. Indeed, within our DFT calculations where we added electrons to NbSe$_2$ to mimic a doping of $\delta \sim 0.2$, we found the amplitude of the CDW decreases by a few tens of %, but does not disappear. An independent test of our scenario is provided by the quantitative analysis of the visible full width at half maximum (*W*=FWHM, Figure 7d). Indeed, if the finite size of the phase domains was the *only* source of superlattice peaks' broadening, we would expect this parameter to be proportional to $1/l \propto \sqrt{\delta}$ (black line in Figure 7d). Indeed, this expression describes the experiment rather well, even though there are clearly additional, weaker mechanisms that are responsible for the finite width at $\delta = 0$. The gradual reduction of the size 3x3 domains is experimentally shown in **Figure S7**.

Figure 7c also shows that maximum intensity of the superlattice peaks. Assuming the integrated intensity is independent of $\delta$ (the reduction due to electron doping is too small), this parameter should behave approximately as $I_{3x3}/I_{1x1} \propto W^{-2}$. In reality the strong background introduces considerable noise around this average expression, yet it does capture the overall trends.

## 6. Conclusion

In summary, we were able to grow high-quality monolayers of Nb$_{1-\delta}$Mo$_\delta$Se$_2$ across the entire alloy composition range ($0 < \delta < 1$), thus bridging two electronically distinct 2D TMD materials. This aliovalent alloy allowed us to investigate the atomic-scale evolution of the electronic ground state across three different phase transitions, i.e., superconductor-metal, CDW, and metal-semiconductor. Our measurements reveal a remarkable robustness of the collective electronic states (CDW and SC) against disorder, and a curious non-monotonic evolution of the SC state with disorder, which strengthens at small Mo concentrations. While some aspects of the evolution of electronic and superconducting properties are seemingly unexpected, and even counterintuitive, our first-principles calculations yield complete microscopic insight into these changes as a function of doping. Our work provides a general methodology to study a wide variety of disorder-driven 2D electronic phase transitions, which remain largely unexplored due to the lack of suitable platforms.

**Experimental Section/Methods**

*Bilayer graphene substrate preparation*: BLG substrates were prepared by annealing 6H-SiC (0001) at 1675 K for 35 minutes in our home-made MBE system operated in ultra-high-vacuum (UHV) conditions (base pressure of ~ $5.0 \times 10^{-10}$ mbar). We used Si-faced polished SiC wafers with resistivity ~ 120 $\Omega \cdot$cm. Before vacuum graphitization of 6H-SiC(0001), they were first degassed in UHV at 920 K for 15minutes and then directly annealed to the target temperature with a heating rate of 10 K • s$^{-1}$.

*Growth of $Mo_\delta Nb_{1-\delta}Se_2$ alloys*: Single-layer $Nb_{1-\delta}Mo_\delta Se_2$ alloys were grown on previously prepared BLG/SiC(0001) by molecular beam epitaxy (MBE). High purity Nb, Mo (99.95%) and Se (99.999%) were simultaneously evaporated from two electron-beam evaporators (Nb and Mo) and a standard Knudsen cell (Se). The evaporation flux ratio between the transition metal elements (Nb and Mo) and Se was kept around 1:30. During the growth the substrate temperature was kept at 820 ± 20 K, and after growth the sample was further annealed in Se environment for 2 minutes. The growth process was monitored by in-situ RHEED (Figure S1(a)) and the growth rate was ~35 minute/monolayer. In order to control the stoichiometry $\delta$ of $Nb_{1-\delta}Mo_\delta Se_2$, we kept the flux of the dominant TM (Nb for $\delta < 0.5$ and Mo for $\delta > 0.5$) constant and proportionally change the flux of the other TM element. After the growth of $Nb_{1-\delta}Mo_\delta Se_2$ alloys, a Se capping layer with a thickness of ~ 5 nm was deposited on the sample surface to protect the film from contamination during transport through air to the UHV-STM chamber. The Se capping layer was subsequently removed prior to the STM experiments by annealing the sample up to 575 K in UHV conditions for 30 minutes. Typical large-scale morphology of single-layer $Nb_{1-\delta}Mo_\delta Se_2$ alloys with relatively high coverage (70%, 25mins growth) is shown in the STM image of Figure S1(b).

*Stoichiometry determination of the $Nb_{1-\delta}Mo_\delta Se_2$ alloys*: The possibility to identify individual Mo (Nb) dopants in atomically resolved STM images of the 2D alloys enabled the nearly exact determination of their stoichiometry for $\delta \leq 0.12$ ($\delta \geq 0.9$). Within these dilute regimes, we found a linear relationship between $\delta$ and the dopant flux ratio, as shown in Figure S2a. Therefore, for non-dilute alloys (0.12 $\leq \delta \leq$ 0.9) where the counting of individual dopants is not possible, their composition was estimated by extrapolation.

*STM/STS measurements*: STM/STS experiments were carried out in a commercial Unisoku UHV, low-temperature and high magnetic field (STM USM-1300) operated in the temperature range T = 0.3 – 4.2 K. STS measurements (Pt/Ir tips) were performed using the lock-in technique with typical a.c. modulations of ~ 30 μV at 833 Hz for the low-bias (~mV) spectra and ~ 5 mV at 833 Hz for the large-bias (~V) spectra. STM/STS data were analyzed and rendered using WSxM software[54].

*First-principles calculations*: Our calculations are based on density functional theory within the projector-augmented wave method[55] as implemented in the VASP code[56,57] using the generalized gradient approximation defined by the Perdew-Burke-Ernzerhof (PBE) functional.[58] We use the following PAWs; Nb $4s^1$, $4p^6$, $4d^4$ electrons, Mo $5s^1$, $5p^6$, $5d^5$ electrons, and Se $4s^2$, $4p^4$ electrons are treated as valence. All calculations use a plane-wave energy cutoff of 400 eV. We use a ($28 \times 28 \times 1$) Γ-centered k-point grid for the monolayer structure when performing structural optimization and calculating the electronic structure. The cell shape and atomic positions of each structure was optimized using a force convergence criteria of 5 meV/Å. All of our calculations using the GGA with spin-orbit coupling (except where stated otherwise).

For the defect calculations we use a ($8 \times 8 \times 1$) supercell of SL-NbSe$_2$. We consider different configurations (see main text) and for each structure we relax all of the atomic coordinates and determine the total energy. The formation energy, $E^f$, of for example, Mo$_{Nb}$ is defined as:

$$E^f = E_{\text{tot}}(\text{Mo}_{\text{Nb}}) - E_{\text{tot}}(\text{NbSe}_2) - \mu_{\text{Mo}} + \mu_{\text{Nb}}$$

where $E_{\text{tot}}(\text{Mo}_{\text{Nb}})$ is the total energy of the supercell with Mo substituting for the Nb site, $E_{\text{tot}}(\text{NbSe}_2)$ is the total energy of the pristine NbSe$_2$ supercell, and $\mu_{\text{Mo}}$ and $\mu_{\text{Nb}}$ are chemical potentials for Mo and Nb respectively. The limiting phase for $\mu_{\text{Mo}}$ is the formation enthalpy of monolayer MoSe$_2$. All of the defect calculations are performed with the (3x3x1) *k*-point grid.

For the considerations of doping beyond the dilute limit we consider two different approaches; the virtual crystal approximation (VCA) which uses unit cell calculations and explicit supercell calculations using an orthorhombic approximately square supercell with 30 NbSe$_2$ formula units that is constructed from the unit cell of the ML structure. The k-point grid for structural relaxation of each supercell is scaled with respect to the ($28 \times 28 \times 1$) Γ-centered k-point grid we use for calculations of the unit cell. For the VCA and supercell calculations we kept the structure fixed at the NbSe$_2$ lattice parameters since our focus is on doping with low concentrations of Mo.

To determine the ferromagnetic spin susceptibility, we used collinear fixed-spin moment (FSM) calculations (sometimes referred to as the constrained local moments approach). In our collinear FSM calculations, we constrain the magnitude of the magnetic moment on the Nb atom. Performing these calculations allows us to determine the change in energy with respect to the non-magnetic ground state as a function of the total magnetization, *m*. We then fit our results to an expansion of the total energy as a function of *m*. The ferromagnetic spin susceptibility obtained from FSM calculations (see main text) is sensitive to the choice in energy convergence threshold, and the number of magnetization values used in the fit to expansion in the total energy as a function of magnetic moment. We use an energy convergence threshold of $10^{-8}$ eV, and up to 50 magnetization versus energy points between 0 $\mu_B$ and 0.6 $\mu_B$ for all of the FSM calculations.

The fixed spin moment calculations and the density of states of the alloys are obtained using VCA calculations for Mo content that corresponds to $\delta$ = 0, 0.01, 0.02, 0.05, 0.1, 0.15 and 0.2. The VCA calculations use the same k-point grid as the unit cell calculations. To verify our VCA calculations we also performed FSM calculations using supercell calculations for $\delta = 0.033$ and $\delta = 0.066$ and find semi-quantitative agreement between the VCA and supercell calculations.

**Supporting Information**

Supporting Information is available from the Wiley Online Library or from the author.

**Acknowledgements**

We acknowledge fruitful discussions with Félix Ynduráin. M.M.U. acknowledges support by the ERC Starting grant LINKSPM (Grant 758558), by the Spanish MINECO under grant no. PID2020-116619GB-C21 and by the Gobierno Vasco under grant PIBA-2020-1-0036. R.H. acknowledges support from Marie Skołodowska-Curie Individual Fellowships under HORIZON 2020 program for project MAGTMD (101033538). D.W was supported by the Office of Naval Research (ONR) through the Naval Research Laboratory's Basic Research Program. I.I.M. was supported by ONR through grant N00014-20-1-2345.

Supporting Information

# Nontrivial doping evolution of electronic properties in Ising-superconducting alloys

*Wen Wan, Darshana Wickramaratne, Paul Dreher, Rishav Harsh, I. I. Mazin, and Miguel M. Ugeda\**

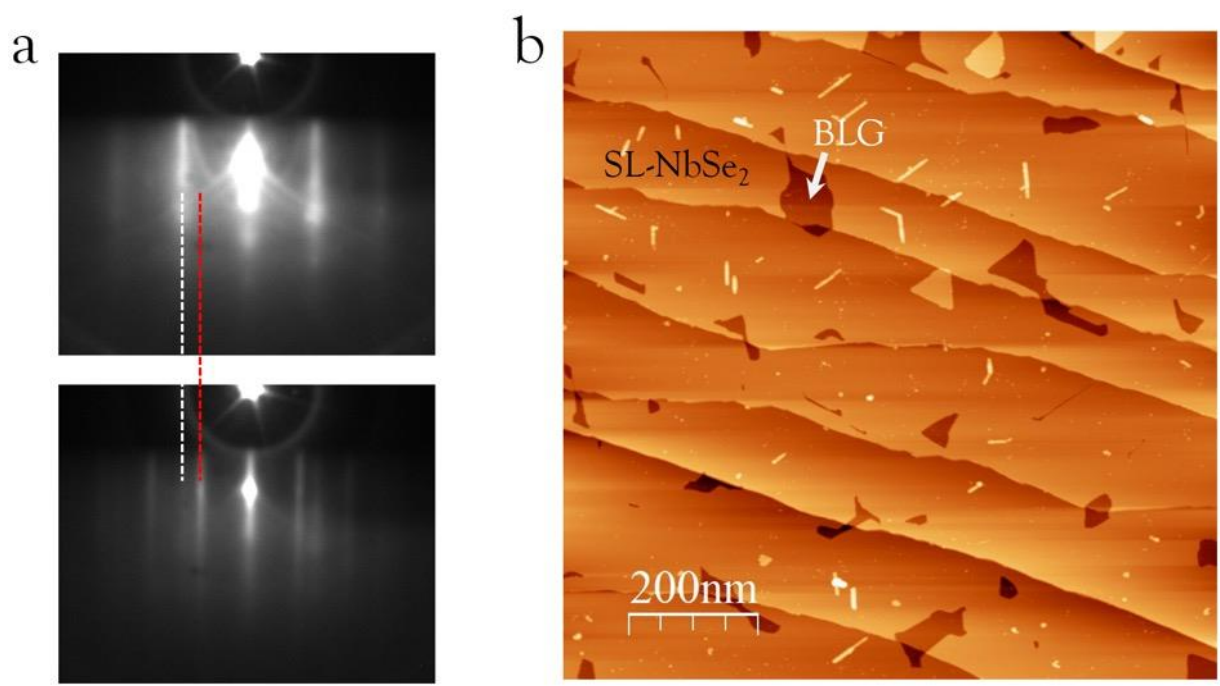

**Figure S1.** Morphology of single-layer $Nb_{1-\delta}Mo_\delta Se_2$. a) RHEED pattern before (upper) and after (lower) the growth of a $Nb_{1-\delta}Mo_\delta Se_2$ alloy on BLG/SiC(0001). BLG and $Mo_\delta Nb_{1-\delta}Se_2$ diffraction patterns are indicated by the white and red dash lines, respectively. b) Typical large-scale STM topography of a $Nb_{0.7}Mo_{0.3}Se_2$ monolayer on BLG/SiC(0001).

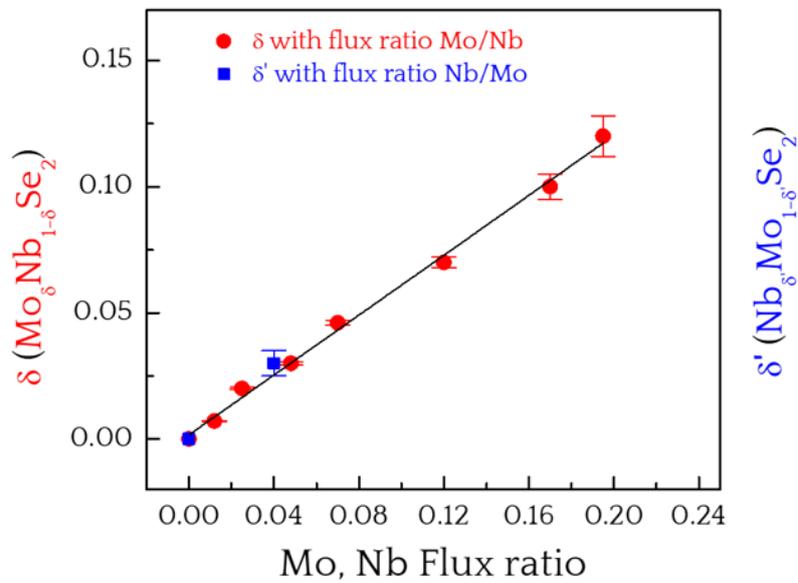

**Figure S2.** Stoichiometry determination. Linear relationship between the stoichiometry δ (δ') and the Mo (Nb) flux ratio for the Nb$_{1-\delta}$Mo$_\delta$Se$_2$ (Nb$_{\delta'}$Mo$_{1-\delta'}$Se$_2$) alloys.

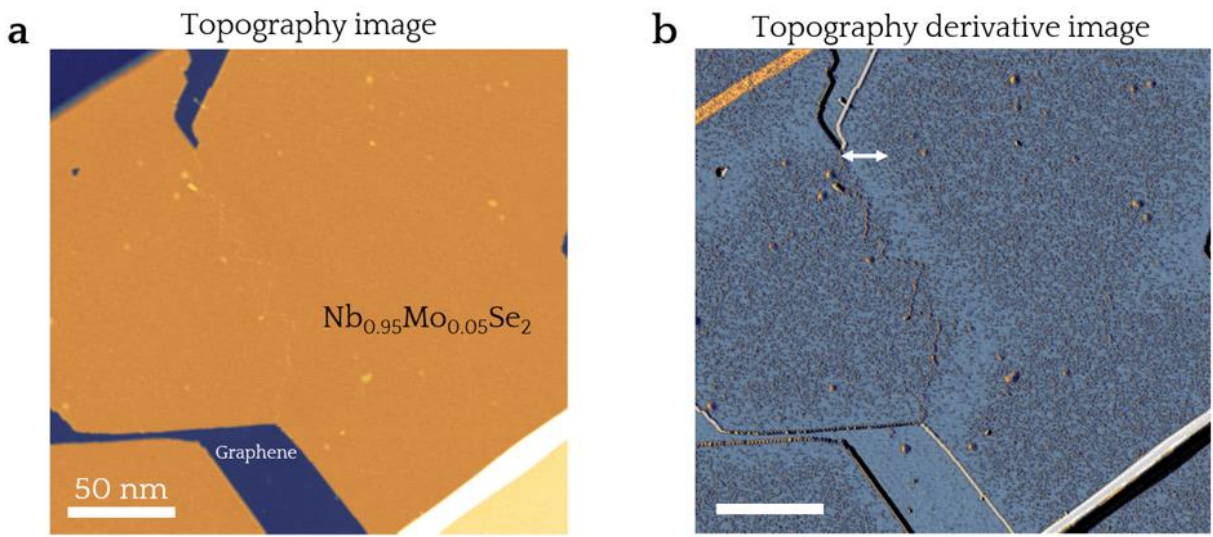

**Figure S3.** Homogeneity of the Mo:doped Nb$_{1-\delta}$Mo$_\delta$Se$_2$ alloy. a) Large-scale STM image of the Nb$_{0.95}$Mo$_{0.05}$Se$_2$ monolayer ($V_s = -0.5$ V, $I_t = 0.01$ nA, $T = 4.2$ K). b) Topography derivative image of a) to enhance the contrast of the Mo dopant concentration, which is homogenous except for a Mo depletion near the edges of the domains (~ 10 nm).

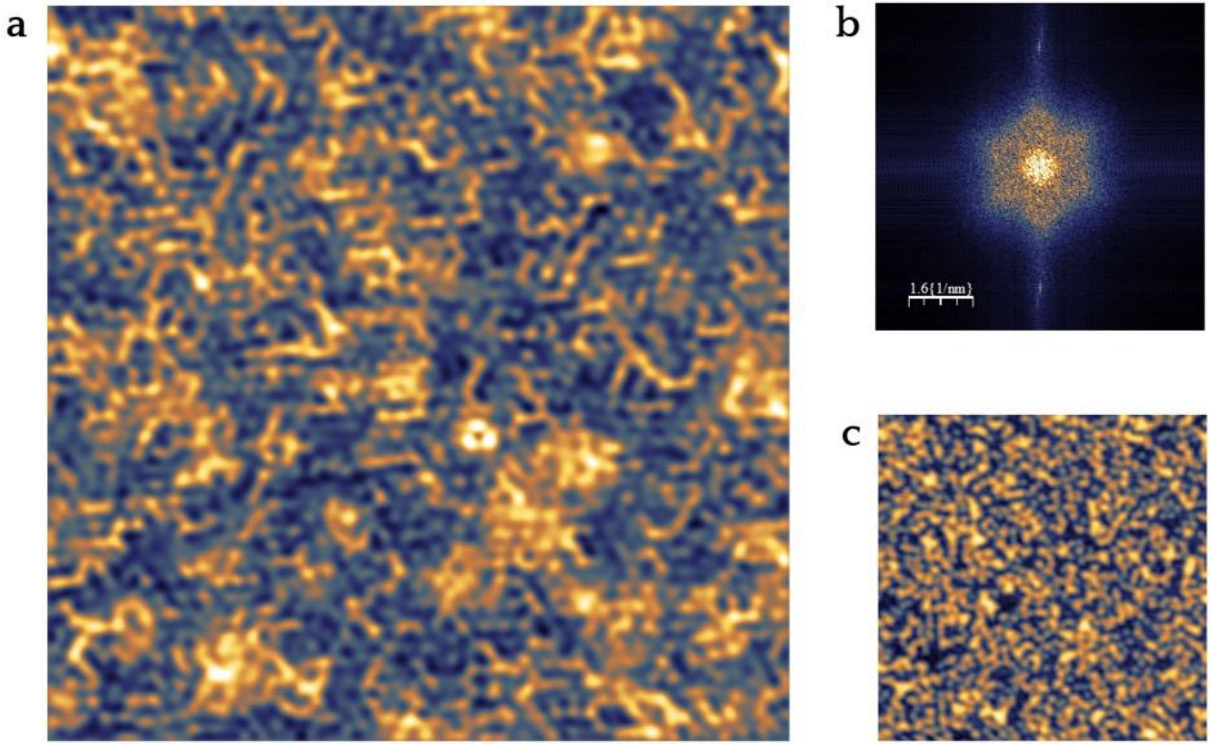

**Figure S4.** Striped substitutional Mo arrangement. a) STM image of the $Nb_{0.8}Mo_{0.2}Se_2$ monolayer showing stripes of Mo atoms ($V_s = + 2$ V, $I_t = 0.15$ nA, T = 4.2 K). b) FFT image of the topography in a) showing the six-fold angular stripe arrangement. c) STM image of the same region as in a) taken at the opposite polarity (occupied states) ($V_s = - 0.4$ V, $I_t = 0.15$ nA, T = 4.2 K).

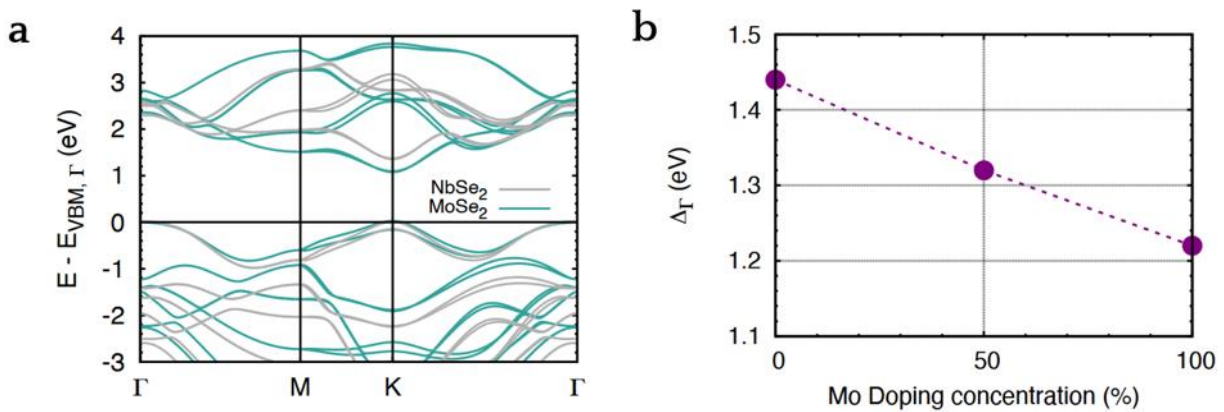

**Figure S5.** Band structure with spin-orbit coupling of the monolayer TMDs. a) Band structure of monolayer $NbSe_2$ (grey) $MoSe_2$ (teal) plotted with respect to the energy of the valence band state at Γ. b) Energy separation, $\Delta_\Gamma$, between the highest occupied state at Γ and the next lowest Se-derived state

as a function of Mo concentration obtained using a 2x1x1 supercell of the hexagonal unit cell of the monolayer TMD.

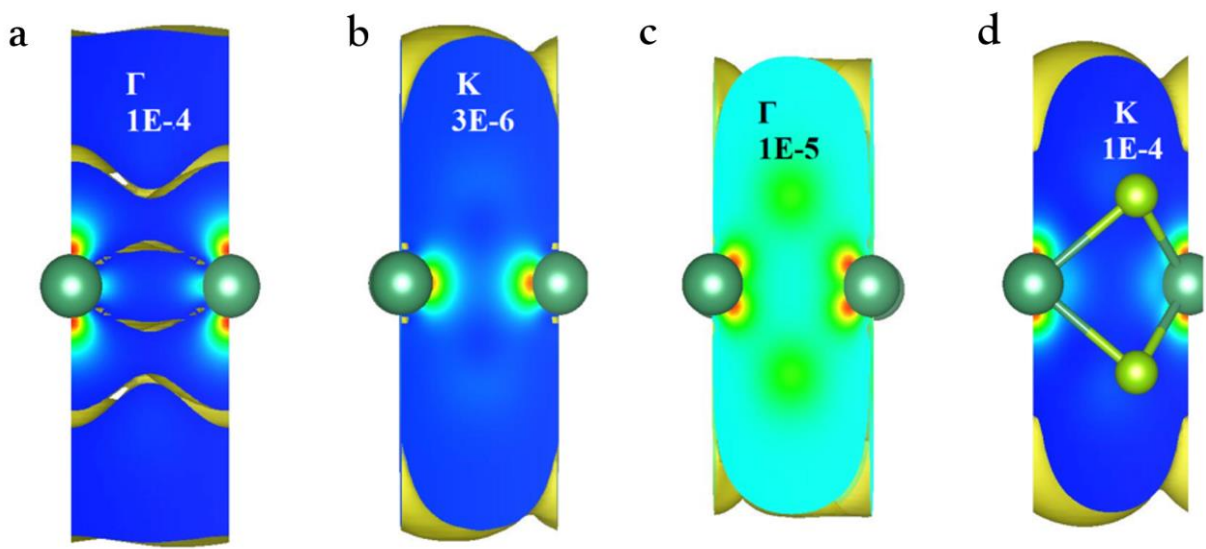

**Figure S6.** Electron density. a) Electron density (in units of $e/A^3$) isosurfaces for the valence band maxima (VBM) at $\Gamma$ a) and K b). Note that at the same distance from the NbSe$_2$ layer the charge density contributed by the states at $\Gamma$ is ~30 times larger than that from K. c) and d) show the electron density isosurfaces for the conduction band minima (CBM). In that case, the density from the states at K is 10 times larger (the color does not carry any useful information, only the spatial extent matters).

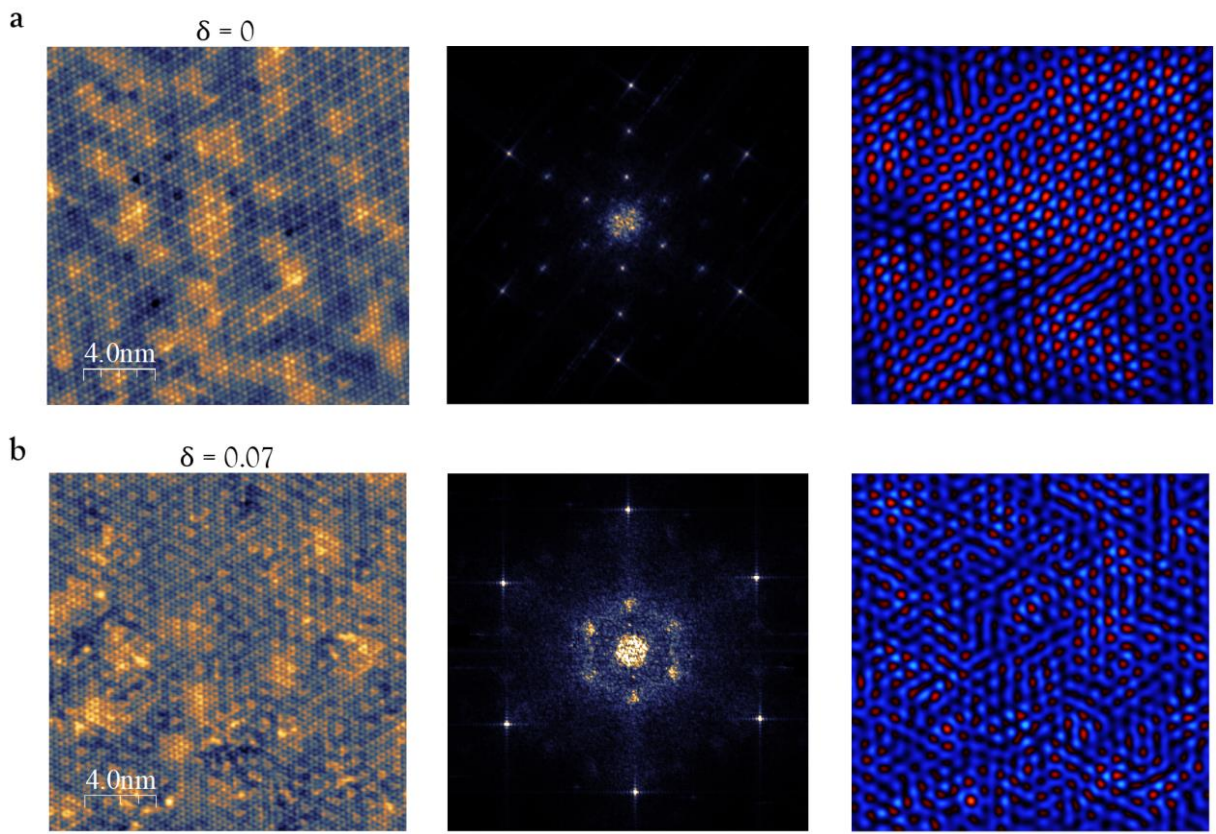

**Figure S7.** 3x3 domains. a) Left, STM Topograph of pristine NbSe$_2$ ($\delta = 0$) and, center, its corresponding FFT. Right, inverse FFT of the integrated 3x3 peaks. Red regions indicate crystalline 3x3 domains and blue regions represent domain boundaries. Parameters: $V_s = -0.09$ V, $I_t = 2$ nA, T = 0.34 K. b) Same as a) for the Nb$_{0.93}$Mo$_{0.07}$Se$_2$ alloy shown in Figure 7a. Parameters: $V_s = +0.05$ V, $I_t = 0.7$ nA, T = 0.34 K.